\documentclass[aps,prd,preprint,superscriptaddress,tightenlines,nofootinbib]{revtex4}

\usepackage{graphicx}
\usepackage{dcolumn}
\usepackage{bm}
\usepackage{ifthen}%

\begin{document}

\preprint{CLNS 05/1938}       
\preprint{CLEO 05-26}         

\title{Experimental Limits on Weak Annihilation Contributions to $b\to u\ell\nu$ Decay}


\author{J.~L.~Rosner}
\affiliation{Enrico Fermi Institute, University of
Chicago, Chicago, Illinois 60637}
\author{N.~E.~Adam}
\author{J.~P.~Alexander}
\author{K.~Berkelman}
\author{D.~G.~Cassel}
\author{J.~E.~Duboscq}
\author{K.~M.~Ecklund}
\author{R.~Ehrlich}
\author{L.~Fields}
\author{L.~Gibbons}
\author{R.~Gray}
\author{S.~W.~Gray}
\author{D.~L.~Hartill}
\author{B.~K.~Heltsley}
\author{D.~Hertz}
\author{C.~D.~Jones}
\author{J.~Kandaswamy}
\author{D.~L.~Kreinick}
\author{V.~E.~Kuznetsov}
\author{H.~Mahlke-Kr\"uger}
\author{T.~O.~Meyer}
\author{P.~U.~E.~Onyisi}
\author{J.~R.~Patterson}
\author{D.~Peterson}
\author{E.~A.~Phillips}
\author{J.~Pivarski}
\author{D.~Riley}
\author{A.~Ryd}
\author{A.~J.~Sadoff}
\author{H.~Schwarthoff}
\author{X.~Shi}
\author{S.~Stroiney}
\author{W.~M.~Sun}
\author{T.~Wilksen}
\author{M.~Weinberger}
\affiliation{Cornell University, Ithaca, New York 14853}
% ...
\collaboration{CLEO Collaboration} 
\noaffiliation


\newcommand{\vub}{$V_{ub}$}
\newcommand{\vcb}{$V_{cb}$}
\newcommand{\modvub}{$|V_{ub}|$}
\newcommand{\modvcb}{$|V_{cb}|$}

\newcommand{\btoulnu}{\protect{\lowercase{b \to u \, \ell \, \nu}}}
\newcommand{\btoclnu}{\protect{\lowercase{b \to c \, \ell \, \nu}}}

\newcommand{\BtoXulnu}{$B \to X_\protect{\lowercase{u}} \, \ell \, \nu$}
\newcommand{\BtoXclnu}{$B \to X_\protect{\lowercase{c}} \, \ell \, \nu$}
\newcommand{\BtoXlnu}{$B \to X \, \ell \, \nu$}

\newcommand{\btou}{$\protect{\lowercase{b \to u}}$}
\newcommand{\btoc}{$\protect{\lowercase{b \to c}}$}

\newcommand{\bbar}{$B\bar{B}$}

\date{January 10, 2006}

\begin{abstract} 
We present the first experimental limits on high-$q^2$ contributions to charmless semileptonic $B$ decays of the form expected from the Weak Annihilation (WA) decay mechanism.  Such contributions could bias determinations of $|V_{ub}|$ from inclusive measurements of $B \rightarrow X_u \ell \nu$.  Using a wide range of models based on available theoretical input we set a limit of 
$\Gamma_{\mathrm{WA}}/\Gamma_{b \rightarrow u}< 7.4$\% (90\% confidence level) on the WA fraction, and assess the impact on previous inclusive determinations of $|V_{ub}|$.
\end{abstract}

\maketitle
A precise determination of the magnitude of the Cabibbo-Kobayashi-Maskawa (CKM) matrix element $V_{ub}$ with well-understood uncertainties is one of the highest priorities in heavy-flavor physics.  Recently, significant experimental progress in the determination of $|V_{ub}|$ through inclusive measurements of semileptonic $B$ decays has been achieved with pioneering work such as the incorporation of $b\to s \gamma$ spectral information for improved modelling \cite{Bornheim:2002du} and the use of large $B$-tagged samples \cite{Aubert:2003zw,Bizjak:2005hn}. Theoretical advances include the quantitative evaluation of the leading and
subleading contributions to the partial $B\to
X_u\ell\nu$ width in restricted regions of phase
space~\cite{Burrell:2003cf,Bosch:2004th,Bosch:2004cb,Hoang:2005pj,Lange:2005yw}.
Several concerns remain, one of which is the
the ``Weak Annihilation'' (WA)
contribution~\cite{Bigi:1993bh,Neubert:1996we,Voloshin:2001xi} to the
total $b\to u\ell\nu$ rate. 

\newboolean{useold}
\setboolean{useold}{false}

\ifthenelse{\boolean{useold}}{
The four-quark WA operator is formally
suppressed at order $(\Lambda/m_b)^3$ in the heavy quark Operator
Product Expansion (OPE),  but  receives an enhancement of
$16\pi^2$ relative to other ${\cal O}(m_b^-3) operators, leading to rate estimates, such as
\cite{Neubert:1996we,Voloshin:2001xi} 
\begin{equation}
  \Gamma_{\rm WA} / \Gamma_{b \to u} \approx 0.03 \left( \frac{f_B}{0.2\;
  {\rm GeV}} \right)^2 \left(\frac{B_2 - B_1}{0.1}\right),
\end{equation}
at the few percent level.  A  non-zero WA contribution requires violation of
QCD factorization, parameterized above by  $B_2-B_1$.
 Little is known about the scale
of this violation since it is fundamentally non-perturbative.  Because WA is expected to be concentrated
in phase space near $q^2 \approx m_b^2$, its \textit{relative} importance is magnified by kinematic requirements that accentuate
the high $q^2$ region to isolate \btou\ from
the large \btoc\ background.  Limiting WA is thus important for understanding the precision of inclusive $|V_{ub}|$ determinations.
}{
The WA contribution arises from a  four-quark operator at order 
$(\Lambda/m_b)^3$ in the heavy quark Operator
Product Expansion (OPE) in which, crudely, the $b$ quark annihilates within the $B$ meson.
Due to an enhancement of
$16\pi^2$ relative to other ${\cal O}(m_b^{-3})$ operators, WA rate estimates,
\cite{Neubert:1996we,Voloshin:2001xi} 
\begin{equation}
  \Gamma_{\rm WA} / \Gamma_{b \to u} \approx 0.03 \left( \frac{f_B}{0.2\;
  {\rm GeV}} \right)^2 \left(\frac{B_2 - B_1}{0.1}\right),
\end{equation}
are a sizeable few percent.  A  non-zero WA contribution requires violation of
QCD factorization, parameterized above by  $B_2-B_1$.
 Little is known about the scale
of this violation since it is fundamentally non-perturbative.  Because WA is expected to be concentrated
in phase space near $q^2 \approx m_b^2$, its \textit{relative} importance is magnified by kinematic requirements that accentuate
the high $q^2$ region to isolate \btou\ from
the large \btoc\ background.  Limiting WA is thus important for understanding the precision of inclusive $|V_{ub}|$ determinations.
}

In this  Letter, we describe a search~\cite{TOMeyer:thesis}  for a $\btoulnu$ rate at large $q^2$, such as would be expected from WA.  We use the
15.5~fb$^{-1}$ of data collected at the $\Upsilon(4S)$ resonance with the CLEO~II \cite{Kubota:1992ww}, CLEO~II.V \cite{Hill:1998ea} and CLEO~III \cite{Peterson:1994cd} detectors at the Cornell Electron Storage
Ring (CESR).  The analysis employs the missing momentum and missing energy
techniques used in analyses of semileptonic moments
\cite{Cronin-Hennessy:2001fk,Mahmood:2002tt} and originally developed
in studies of exclusive charmless semileptonic decay
\cite{Alexander:1996qu,Athanas:1997eg,Athar:2003yg}.  The procedure
takes advantage of the near hermeticity  of
the CLEO detectors to estimate the four momentum of a single missing neutrino
from the missing four momentum $p_{\text{miss}}=(E_{\text{miss}},\vec{p}_{\text{miss}})$  in an event,
where $E_{\text{miss}}\equiv 2E_{\text{beam}}-\sum E_i$ and
$\vec{p}_{\text{miss}}\equiv \vec{p}_{CM}-\sum\vec{p}_i$.

All detector configurations provide acceptance over more than 90\% of the full $4\pi$ solid angle for both charged particles (momentum resolution of 0.6\% at 2 GeV$/c$) and photons (typical $\pi^0$ mass resolution of 6 MeV).  
Charged particles are assigned the most probable mass based on a combination of specific ionization measurements, either time-of-flight or \v Cerenkov radiation angle measurements, and the relative spectra of $\pi^\pm$, $K^\pm$ and $p$ from $B$ decay.  
To optimize the missing-momentum resolution, the selection is optimized to identify a single track for each charged particle originating from the $B$ and $\bar{B}$ decays with high efficiency, and to measure energy deposited in the CsI calorimeter that is not associated with charged tracks or their interaction products.  
The track subset choice is based on event topology rather than individual track quality since high efficiency with minimal double counting is more important than use of only well-measured tracks.
Energy clusters located within 8 cm of tracks determined to project into the CsI calorimeter, or consistent with being ``split off'' from a matched shower, are excluded from the energy and momentum sums.

Electrons satisfying $p>400\text{ MeV}/c$  are identifed over 90\% of $4\pi$ using the
ratio of cluster energy to track momentum in conjunction with specific ionization ($dE/dx$) measurements in the main drift chamber.  Time-of-flight or \v Cerenkov measurements provide additional $e^\pm/K^\pm$ separation in the momentum range with ambiguous $dE/dx$ information.
Particles  in the polar angle range $|\cos\theta|<0.85$ that register hits in counters beyond 5
interaction lengths are
accepted as signal muons. Those with $|\cos\theta|<0.71$ and hits between 3 and 5  interaction lengths  are used in the multiple-lepton veto discussed below.  We restrict signal electrons and muons  to the range $1.5 < p < 3.0 \text{ GeV}/c$.  Within these fiducial and momentum regions
the selection efficiency exceeds 90\%. The probability of misidentifying a charged hadron as  an electron (muon) is about 0.1\% (1\%).

In events with multiple undetected particles,  $p_{\text{miss}}$ represents $p_\nu$ poorly and causes reconstructed
variables in $B\to X_c\ell\nu$ decays to smear beyond their kinematic limits into the
regions of sensitivity for the much rarer $B\to X_u\ell\nu$ process (including WA).  This mechanism provides the dominant background contribution in this analysis. Therefore, we reject events with multiple
identified leptons, which are usually accompanied by  multiple neutrinos.
We also reject events with a net charge not equal to zero, or where 
$\vec{p}_{\text{miss}}$ is consistent with particles lost down the beam pipe
($|\cos\theta_{\text{miss}}|>0.9$).

The missing mass squared $M_{\text{miss}}^2 = E_{\text{miss}}^2 -
|\vec{p}_{\text{miss}}|^2$, which should be zero (within resolution) if only a
sole neutrino is missing, provides further background suppression.
Because the resolution on $E_{\text{miss}}$ is about 60\% larger than
that on $|\vec{p}_{\text{miss}}|$, the $M_{\text{miss}}^2$ resolution $\sigma_{M_{\text{miss}}^2}\approx 2 E_{\text{miss}}\sigma_{E_{\text{miss}}}$.
Requiring $M_{\text{miss}}^2/2E_{\text{miss}}< 0.2
\text{ GeV}$ provides a zero mass requirement at roughly constant
$E_{\text{miss}}$ resolution that enhances the signal relative to the background by over a factor of  two.  We then take $p_\nu =
(E_\nu,\vec{p}_\nu)=(|\vec{p}_{\text{miss}}|,\vec{p}_{\text{miss}})$
in other kinematic calculations.  In particular, we can calculate
$q^2$, the square of the hadronic momentum transfer in semileptonic
decay, via $q^2 = (p_\ell + p_\nu)^2$.  The core resolution on $q^2$
is about $0.6\ \text{GeV}^2$, with a broad high-side tail from
events with more than one undetected particle. 

Finally, we  ensure that the event is consistent with
$e^+e^-\to B\bar{B}$ decay.  We begin with the standard CLEO hadronic
sample, defined by events with at least six primary tracks and a
visible energy of at least 20\% of the center of mass energy.  We
suppress continuum $e^+e^-\to q\bar{q}$ backgrounds and $\tau^+\tau^-$
backgrounds using the ratio ($R_2=H_2/H_0$) of the second to zeroth
Fox-Wolfram moments \cite{Fox:1978vu} and a sphericity-like variable
\cite{Bjorken:1969wi} that is sensitive to momentum flow
perpendicular to the lepton, which is small for continuum
events.  Of the events that satisfy  all other criteria and $q^2>2\,\textrm{GeV}^2$, our continuum suppression rejects 72\% of continuum while
retaining over 90\% of semileptonic $B$ decays.

\begin{figure}
  \begin{center}
    \includegraphics[scale=0.5]{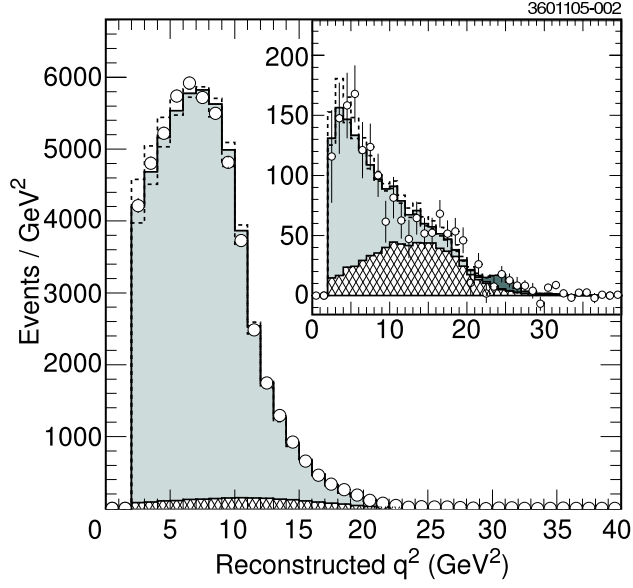}
  \end{center}
  \caption{The continuum- and fake-lepton-subtracted $q^2$ spectra (points) for $p_\ell > 1.5$~GeV/$c$ and $p_\ell > 2.2$~GeV/$c$ (inset) with components $B \rightarrow X_c \ell \nu$ (light grey), $B \rightarrow X_u \ell \nu$ (hatch) and  WA (with $\langle M_x\rangle = 0.293$~GeV, dark grey).  The dashed envelope results from systematic variation of $B \rightarrow X_c \ell \nu$.
    \label{fig:fitresults}}
\end{figure}

We search for evidence of WA or other sources of \mbox{high-$q^2$} $B \rightarrow X_u \ell \nu$ decays by fitting our measured $q^2$ spectra in three lepton-momentum bins.  The lowest bin ($1.5 < p_\ell \leq 2.0$ GeV/$c$) is dominated by $B \rightarrow X_c \ell \nu$ and serves to normalize this background.  The total $B \rightarrow X_u \ell \nu$ yield is determined by the middle ($2.0 < p_\ell \leq 2.2$~GeV/$c$) and highest ($p_\ell > 2.2$~GeV/$c$) bins, while the greatest sensitivity for WA-like processes is provided by the highest momentum bin.  Our $q^2$ spectra for the full sample and for the highest momentum bin are shown in Fig.~\ref{fig:fitresults}.  Corrections have been applied for continuum background (using data collected just below $B \bar{B}$ threshold) and for events in which hadrons were misidentified as signal leptons (using data obtained with a lepton veto folded with measured misidentification probabilities).


\begingroup
\squeezetable
\begin{table*}
  \centering 
  \caption{Summary of ``impact ratios'' for some WA models considered. }
  \label{tab:results}
\begin{tabular}{ccc@{\extracolsep{1em}}cccc}
\hline\hline
$x_0$ (GeV) & $\Lambda$ (GeV)  & $\langle M_X\rangle$ (GeV) & $R_{\text{total}}$ (\%) &  $R_{\text{endpoint}}$ (\%) & $R_{q^2M_x}$ (\%) & $R_{M_x}$ (\%) \\ \hline
0.30 & 0.01 & 0.293 & $1.73\pm0.68\pm0.60$ & $8.24\pm3.04\pm2.34$ & $3.89\pm1.50\pm1.18$ & $2.24\pm0.88\pm0.67$ \\ 
0.30 & 0.05 & 0.328 & $1.40\pm0.69\pm0.58$ & $6.73\pm3.11\pm2.36$ & $3.17\pm1.52\pm1.18$ & $1.82\pm0.89\pm0.66$ \\ 
0.30 & 0.20 & 0.476 & $1.36\pm0.89\pm0.82$ & $5.61\pm3.52\pm3.13$ & $3.06\pm1.97\pm1.77$ & $1.76\pm1.15\pm1.00$ \\ 
0.30 & 0.30 & 0.574 & $1.49\pm1.25\pm1.16$ & $5.19\pm4.16\pm3.82$ & $3.33\pm2.72\pm2.50$ & $1.91\pm1.58\pm1.43$ \\ 
0.30 & 0.50 & 0.773 & $2.95\pm1.89\pm1.89$ & $7.23\pm4.43\pm4.27$ & $6.02\pm3.74\pm3.64$ & $3.51\pm2.24\pm2.13$ \\ 
\hline
0.40 & 0.01 & 0.342 & $1.17\pm0.65\pm0.55$ & $5.62\pm2.98\pm2.28$ & $2.64\pm1.45\pm1.13$ & $1.52\pm0.84\pm0.63$ \\ 
0.40 & 0.05 & 0.369 & $1.07\pm0.67\pm0.58$ & $5.08\pm3.06\pm2.51$ & $2.42\pm1.50\pm1.24$ & $1.38\pm0.87\pm0.70$ \\ 
0.40 & 0.20 & 0.498 & $1.22\pm1.00\pm0.96$ & $4.85\pm3.82\pm3.64$ & $2.76\pm2.22\pm2.11$ & $1.58\pm1.29\pm1.21$ \\ 
0.40 & 0.30 & 0.593 & $1.88\pm1.27\pm1.30$ & $6.18\pm3.99\pm4.01$ & $4.16\pm2.74\pm2.76$ & $2.39\pm1.61\pm1.58$ \\ 
\hline
0.50 & 0.01 & 0.392 & $0.93\pm0.71\pm0.61$ & $4.35\pm3.20\pm2.70$ & $2.10\pm1.58\pm1.34$ & $1.20\pm0.91\pm0.76$ \\ 
0.50 & 0.05 & 0.416 & $0.95\pm0.76\pm0.75$ & $4.36\pm3.34\pm3.44$ & $2.17\pm1.69\pm1.72$ & $1.24\pm0.98\pm0.99$ \\ 
0.50 & 0.10 & 0.452 & $1.03\pm0.84\pm0.94$ & $4.44\pm3.48\pm4.11$ & $2.34\pm1.87\pm2.16$ & $1.34\pm1.08\pm1.26$ \\ 
0.50 & 0.20 & 0.534 & $1.44\pm1.07\pm1.07$ & $5.27\pm3.76\pm3.73$ & $3.25\pm2.37\pm2.34$ & $1.86\pm1.38\pm1.34$ \\ 
0.50 & 0.30 & 0.621 & $1.58\pm1.36\pm1.32$ & $4.90\pm4.07\pm3.88$ & $3.51\pm2.96\pm2.83$ & $2.01\pm1.72\pm1.62$ \\ 
0.50 & 0.50 & 0.806 & $3.42\pm2.32\pm2.35$ & $7.56\pm4.91\pm4.79$ & $6.85\pm4.49\pm4.42$ & $4.04\pm2.72\pm2.64$ \\ 
\hline
0.60 & 0.01 & 0.442 & $0.82\pm0.80\pm0.81$ & $3.64\pm3.45\pm3.55$ & $1.87\pm1.81\pm1.84$ & $1.07\pm1.04\pm1.06$ \\ 
0.60 & 0.05 & 0.465 & $1.01\pm0.87\pm0.91$ & $4.28\pm3.58\pm3.84$ & $2.29\pm1.96\pm2.07$ & $1.31\pm1.13\pm1.19$ \\ 
0.60 & 0.30 & 0.660 & $2.67\pm1.64\pm1.62$ & $7.41\pm4.34\pm4.10$ & $5.82\pm3.47\pm3.32$ & $3.38\pm2.06\pm1.92$ \\ 
0.60 & 0.50 & 0.836 & $4.27\pm2.61\pm2.75$ & $8.71\pm5.07\pm5.10$ & $8.44\pm4.93\pm5.02$ & $5.00\pm3.03\pm3.03$ \\ 
\hline
0.75 & 0.01 & 0.518 & $1.32\pm1.04\pm1.10$ & $5.01\pm3.79\pm4.01$ & $2.99\pm2.31\pm2.43$ & $1.72\pm1.34\pm1.40$ \\ 
0.75 & 0.20 & 0.641 & $3.33\pm1.79\pm1.89$ & $9.20\pm4.65\pm4.70$ & $7.31\pm3.77\pm3.85$ & $4.27\pm2.27\pm2.24$ \\ 
0.75 & 0.30 & 0.719 & $3.43\pm2.09\pm2.09$ & $8.11\pm4.71\pm4.45$ & $7.38\pm4.32\pm4.13$ & $4.31\pm2.61\pm2.43$ \\ 
0.75 & 0.50 & 0.886 & $4.63\pm2.86\pm3.80$ & $8.34\pm4.94\pm6.37$ & $8.97\pm5.28\pm6.85$ & $5.34\pm3.27\pm4.20$ \\ 
\hline
\end{tabular}
\end{table*}
\endgroup

\begin{figure}[b]
  \begin{center}
    \includegraphics[scale=0.5]{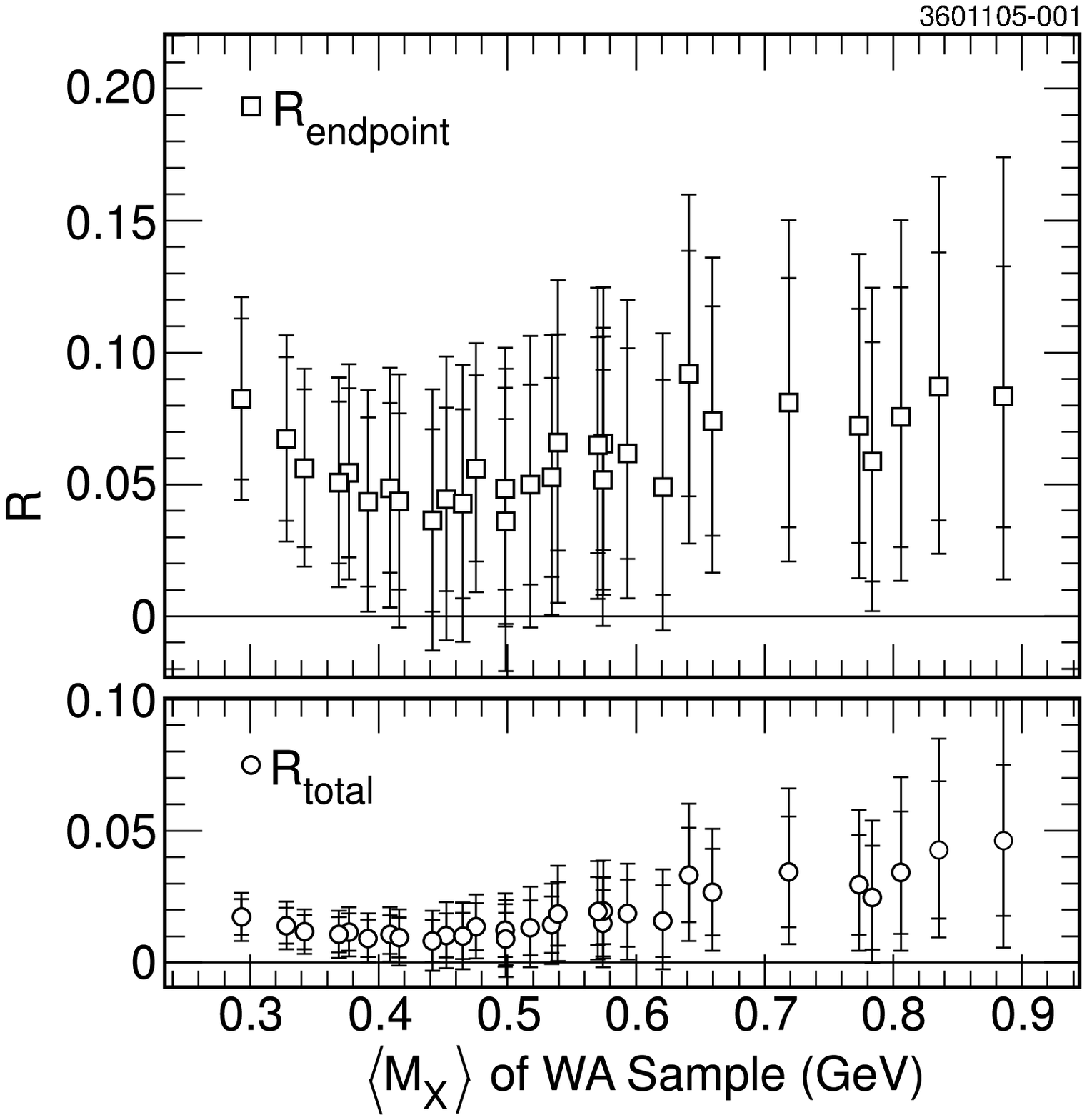}
  \end{center}
  \caption{Fractional size of the WA component for the full phase space (bottom) and restricted to
    $p_\ell>2.2 \textrm{GeV}/c$ endpoint region (top). The statistical (total)
    uncertainties are represented by the inner (full) error bar.  
    \label{fig:impact}}
\end{figure}

The corrected $q^2$ spectra are fitted with $B \rightarrow X_c \ell \nu$ and  $B \rightarrow X_u \ell \nu$ components obtained from simulation. 
The $B \rightarrow X_c \ell \nu$  simulation incorporates all available $B$-decay data, including measured semileptonic decay form factors.  
The $B \rightarrow X_u \ell \nu$ simulation is based on (i) a hybrid model that combines the HQET-based approach of DeFazio and Neubert [24] with known exclusive resonances, and (ii) a simple model for WA that reflects both the kinematics implied by
 $d\Gamma_{\rm WA}/dq^2 \sim \delta(q^2 - m_b^2)$ and  the
intuitive picture where the ``valence'' quarks in the $B$ meson
annihilate and a soft non-perturbative hadronic system $X_u$
materializes.  In this formulation, the lepton-neutrino
pair carries most of the energy ($q^2 \sim M_B^2$), while the hadronic
system has kinematics at the non-perturbative scale $\Lambda_{\rm
  QCD}$.  To describe the spectra of that soft hadronic system, our implementation introduces a probability density function (pdf)
 that is flat out to a cutoff $x_0$, where
an exponential roll-off with slope $\Lambda$ begins. The mass $M_X$ and 
momentum of the hadronic system for a WA decay are drawn
independently from this pdf, uniquely determining the kinematics. The system is
then hadronized into at least two particles or resonances.  The
kinematics of the $\ell\nu$ pair are calculated assuming the $V-A$
structure of the weak current and spin $s=0$ for the hadronic
system. We examine combinations of five $x_0$ and six $\Lambda$ values,
for a total of 30 different WA cases that span a wide range of kinematics.

We perform a separate $\chi^2$ fit for each WA case, with the \btoc, \btou, and WA rates floating independently.  The WA rate is not constrained to be positive.  Acceptable fits to the $q^2$ distributions are obtained for all cases. The $B \rightarrow X_c \ell \nu$ and non-WA $B \rightarrow X_u \ell \nu$ components dominate, and in no case is the WA component more than two standard deviations above zero (combined statistical and systematic). The most significant WA yield  is obtained for the case with $\langle M_x\rangle=0.293$~GeV (shown in Fig.~\ref{fig:fitresults}), and appears to result from an overlap with a downward fluctuation in the $q^2$ distribution of the sample used for continuum subtraction. 

From each
fit's results for the WA and  non-WA \btou\ rates,
the ratio $R \equiv \Gamma_{\rm WA}/\Gamma_{b \to u}$ is computed for 
the full phase space (total) and for restricted phase space regions that have been
used in previous
inclusive \modvub\ measurements: $p_\ell > 2.2$~GeV$/c$ (endpoint);
 $p_\ell > 1.0\,\textrm{GeV}/c$, $q^2 > 8.0\,\textrm{GeV}^2$ and $M_X < 1.7$~GeV ($q^2M_x$);
 and $p_\ell > 1.0\,\textrm{GeV}/c$,  $M_X < 1.55$~GeV ($M_x$).
These ratios constrain the extent to which a measured rate can be biased away from current
theoretical estimates because of a localized WA contribution. 
The ratios for the full-phase-space and endpoint cases are shown in Fig.~\ref{fig:impact} and the results for a subset of the cases considered are summarized in Table~\ref{tab:results}.
In each region our results, which are statistics-limited, set non-trivial constraints on a localized WA enhancement.

The primary systematic uncertainties arise from experimental effects
related to reconstruction of the neutrino, such as the absolute $K_L$
and $b\to c\to s\ell\nu$ rates and spectra, the efficiency and
resolution for charged particle and photon detection, modeling and
rejection of charged hadronic showers, and charged hadron
identification~\cite{Athar:2003yg}.   The $B\to X_c\ell\nu$ modeling
systematic estimate includes variations of the branching
fractions at levels commensurate with recent measurements, and variations of form
factors at levels several standard deviations from
recent average results \cite{Alexander:2005cx}.   The $B\to X_u\ell\nu$ modeling  systematic
includes a variation  of the inclusive shape function similar to Ref.~\cite{Bornheim:2002du}
and variation of the $X_u$ hadronization model.  In the ratios of the
WA component relative to the $B\to X_u\ell\nu$ component, many common
systematics related to luminosity, fake rates, etc.~largely cancel.
Table~\ref{tab:systematics} summarizes the systematic contributions for the 
WA model shown in Fig.~\ref{fig:fitresults}.  Shifts observed with systematic cross checks such
as  floating individual components of the $\btoclnu$ background,
floating individual classes of mistakes (e.g., extra $K_L$ or extra $\nu$) in the
$\btoclnu$ background sample, and even more extreme variations such as eliminating the $D\ell\nu$ and nonresonant $\btoclnu$  components are commensurate with the quoted systematics.

\begin{table}
  \centering 
  \caption{Systematic uncertainties (WA model of Fig.~\protect\ref{fig:fitresults}).}  
  \label{tab:systematics}
\begin{tabular}{lcc}
\hline\hline
Source  & $\Delta R_{\text{tot}}$ & $\Delta R_{\text{tot}}/R_{\text{tot}}$(\%) \\ \hline
      $\gamma$ efficiency        & 0.00177 &  10.2 \\
      tracking efficiency             & 0.00247 & 14.3 \\
      $E_\gamma$ resolution  & 0.00095 &  5.5 \\
      $p_{\rm trk}$ resolution    & 0.00134 &  7.7 \\
      $K_L$ multiplicity              & 0.00013 &  0.8 \\
      hadronic shower modeling   & 0.00118 &  6.8 \\
      hadronic shower veto       & 0.00065 &  3.8 \\
      particle identification         & 0.00078 &  4.5 \\
      $b \to c \to s \ell \nu$         & 0.00020 &  1.1 \\
      $\btoclnu$ modeling         & 0.00349 & 20.1 \\
      $\btoulnu$ modeling         & 0.00309 & 17.9 \\
      \hline
      \textbf{Total}             & \textbf{0.00601} & \textbf{34.7}\\
\hline
\end{tabular}
\end{table}

To limit the bias in rate measurements quantitatively, we parameterize
the variation in the central value and total uncertainty of each WA fraction ($R$)
as a function of $\langle M_X\rangle$.  90\% confidence
limits (CL) are then calculated assuming a 
flat probability distribution in $\langle M_X\rangle$ over the range we have investigated, resulting
in $R_{\text{total}}<7.4\%$, $R_{\text{endpoint}}<15.5\%$, $R_{q^2M_x}<14.5\%$
and $R_{M_x}<8.6\%$.  Limits on a bias of \modvub\ are half these values.
 These results provide the first concrete constraint on one of the three important uncertainties in extraction of \modvub\  for which only more speculative results \cite{Eidelman:2004wy} have existed to date. They also place \modvub\ from endpoint analyses on a much stronger footing, where
the 8\% bound (90\% CL) we find for a bias in an endpoint-based \modvub\   is much more
restrictive than the  estimate $\sigma_{\mathrm{WA}}\approx10-20$\% \cite{Eidelman:2004wy}.  Studies like these are crucial for  inclusive determinations of \modvub\ to achieve a 5\% precision goal robustly (already achieved statistically).

In summary, we have obtained the first experimental limits on the potential bias in inclusive determinations of $|V_{ub}|$  from a localized contribution to the $q^2$ distribution, as could arise from weak annihilation. The method presented here
is one of several, including study of
semileptonic rates in the charm sector and
 of the moments of the $B$ semileptonic 
$q^2$ distribution \cite{Gambino:2005tp}, that will be needed to
understand weak annihilation and its impact upon
\modvub.

We gratefully acknowledge the effort of the CESR staff
in providing us with excellent luminosity and running conditions.  
We thank G. Paz, Z. Ligeti, M. Luke and M. Neubert for helpful discussions. 
This work was supported by 
the A.P.~Sloan Foundation,
the National Science Foundation,
and the U.S. Department of Energy.

\bibliography{WA_prl}
\end{document}